\UseRawInputEncoding

\documentclass[aps,prl,twocolumn,amsmath,amssymb,nofootinbib,superscriptaddress,floatfix]{revtex4-1}

\usepackage{amsmath}
\usepackage{amssymb}
\usepackage{amsthm}
\usepackage{mathtools}
\usepackage{mathdots}
\usepackage{amsfonts}
\usepackage{bbm}
\usepackage{xr}
\usepackage{bbold}
\usepackage{newpxtext}
\usepackage{epsfig,color,dsfont,upgreek,physics}
\usepackage{mathrsfs}
\usepackage{multirow}
\usepackage[makeroom]{cancel}

\usepackage{graphicx}
\usepackage{dcolumn} 
\usepackage{bm} 
\usepackage{float} 
\usepackage[driverfallback=dvipdfm]{hyperref} 
\usepackage{longtable}
\usepackage{ulem}   
\allowdisplaybreaks

\renewcommand\[{\begin{equation}}
\renewcommand\]{\end{equation}}

\begin{document}

\title{Entanglement and Topology in Su-Schrieffer-Heeger Cavity Quantum Electrodynamics}

\author{Daniel Shaffer}
\affiliation
{
Department  of  Physics,  Emory  University,  400 Dowman Drive, Atlanta,  GA  30322,  USA
}

\author{Martin Claassen}
\affiliation
{
Department  of  Physics and Astronomy,  University of Pensylvania, 209 South 33rd Street,  PA  19104,  USA
}

\author{Ajit Srivastava}
\affiliation
{
Department  of  Physics,  Emory  University,  400 Dowman Drive, Atlanta,  GA  30322,  USA
}

\author{Luiz H. Santos}
\affiliation
{
Department  of  Physics,  Emory  University,  400 Dowman Drive, Atlanta,  GA  30322,  USA
}

\begin{abstract}
Cavity materials are a frontier to investigate the role of light-matter interactions on the properties of electronic phases of matter. In this work, we raise a fundamental question: can non-local interactions mediated by cavity photons destabilize a topological electronic phase? We investigate this question by characterizing  entanglement, energy spectrum and correlation functions of the topological Su-Schrieffer-Heeger (SSH) chain interacting with an optical cavity mode. Employing density-matrix renormalization group (DMRG) and exact diagonalization (ED), we demonstrate the stability of the edge state and establish an area law scaling for the ground state entanglement entropy, despite long-range correlations induced by light-matter interactions. These features are linked to gauge invariance and the scaling of virtual photon excitations entangled with matter, effectively computed in a low-dimensional Krylov subspace of the full Hilbert space. This work provides a framework for characterizing novel equilibrium phenomena in topological cavity materials.
\end{abstract}

\date{\today}

\maketitle

\noindent

\noindent
\textit{Introduction--} 
Ever since Purcell's seminal discovery \cite{Purcell_1946} that light-matter interactions (LMI) can be controlled by engineering electromagnetic vacuum, cavity quantum electrodynamics (cQED) \cite{HarocheKleppner1989, Walther_Cavity_2006} has been a fruitful platform to create and manipulate light-matter hybrids. Notable experimental progress in the last decade has enabled ultra-strong coupling regimes where LMI is comparable to or even more significant than the bare cavity and matter excitation energy scales, \cite{Garcia-Vidal_Manipulating_2021, Hübener2021, FriskKockum2019, Forn-Diaz_RMP_2019} opening a promising path to alter the equilibrium properties of quantum materials with quantum light.
\cite{Schlawin_Cavity-Mediated_2019,allocca2019cavity, Ashida_Quantum_2020,latini2021ferroelectric,jarc2022cavity}

Quantum entanglement is inherently part of the description of strongly interacting light-matter systems, for LMI entangles photons and charged particles, resulting in hybrid many-body states containing virtual excitations.\cite{Ciuti_QuantumVacuum_2005} Nevertheless, the nature of quantum entanglement in the regime where light strongly interacts with many-body electronic systems remains to be harnessed.
In particular, in contrast with the pivotal role played by entanglement as a universal marker of long-range entangled topological order\cite{hamma2005ground, kitaev-preskill-2006, levin-wen-2006, Li_Haldane_2008} and short-range entangled symmetry-protected topological states \cite{pollmann2010entanglement, Fidkowski_Entanglement_2010,turner2011topological, chen2011classification, schuch2011classifying, Santos_RK_2015}, the scaling regimes of entanglement in topological matter interacting with cavity fields remains poorly understood. This issue is central to a potential classification of 
hybrid light-matter phases, as well as a timely endeavor given the observed breakdown of topological protection in quantum Hall systems \cite{Appugliese_Breakdown2022} strongly interacting with THz cavity modes.

\begin{figure}
    \includegraphics[width = 7.0 cm]{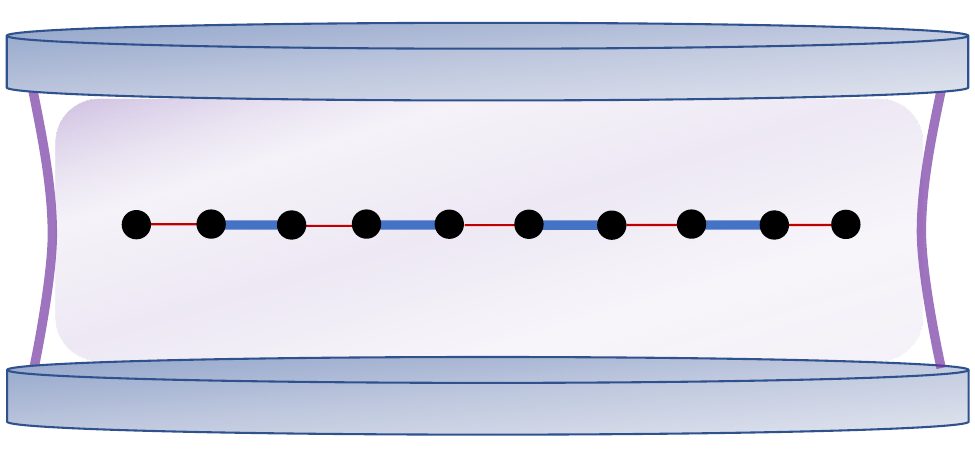}
    \caption{Fermionic Su-Schrieffer-Heeger chain interacting with an optical cavity mode (purple). Intra- and inter-hopping amplitudes $t_e$ and $t_o$ respectively represented by red and blue bonds.}
    \label{fig: SSH-cavity}
\end{figure}

In this Letter, we investigate the one-dimensional (1D) Su-Schrieffer-Heeger (SSH) spinless fermionic chain \cite{SSH_PRL_1979} coupled to a single optical mode as a paradigm to address the effects of LMI onto equilibrium properties of topological fermionic matter. At half-filling, the SSH chain has a trivial and a topological gapped phase separated by a phase transition upon tuning the intra- and inter-unit cell hopping amplitudes, as shown in Fig.\ref{fig: SSH-cavity}. While both phases display similar area law scaling of the ground state entanglement entropy (EE), the low eigenvalues of the entanglement spectrum are two-fold degenerate due to the presence of non-trivial edge states \cite{pollmann2010entanglement,Fidkowski_Entanglement_2010,turner2011topological}. 
In the 
SSH-cQED system
with electrons strongly 
interacting with a single photonic mode, a burning question is to characterize the role of photon-mediated non-local interactions on the system's short-range entanglement and topological properties. 

\begin{figure*}[t]
    \centering
    \includegraphics[width=0.99\linewidth]{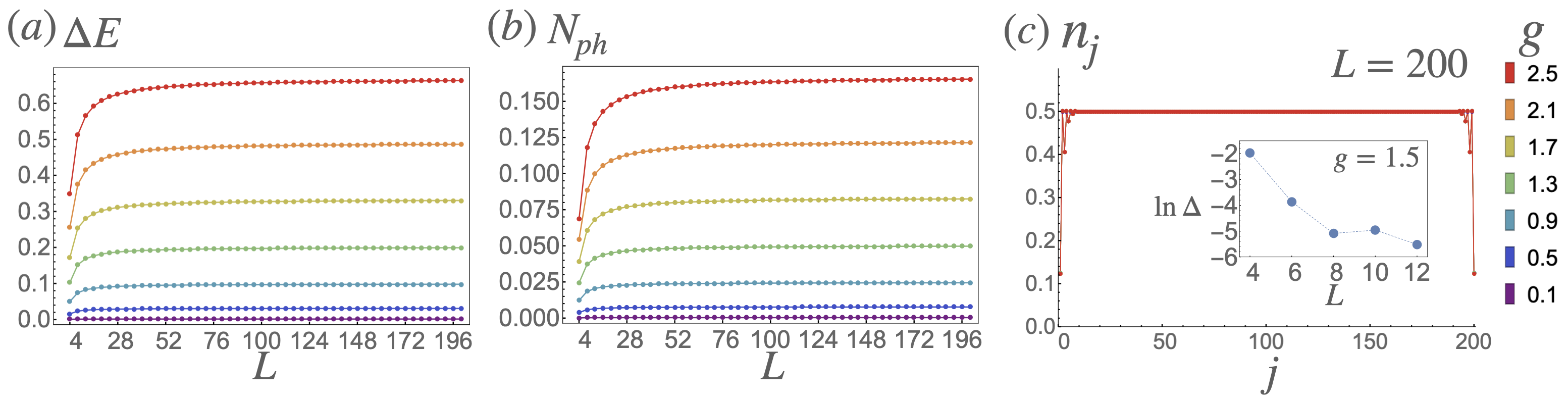}
    \caption{DMRG simulation of cavity SSH system for light-matter coupling $g \in [0.1, 2.5]$
    (color coding legend on the right) for system sizes incremented in steps of \(4\) up to \(L=200\). (a) Change in ground state energy in the presence of light-matter coupling in the dimerized limit (\(t_e=0\)), \(\Delta E= E(g)-E(0)\).  Diamagnetic response ($\Delta E > 0$) saturates to a finite value for large $L$. (b) Average number of photons \(N_{ph}=\langle a^\dagger a\rangle\) in presence of light-matter coupling in the dimerized limit (\(t_e=0\)). \(N_{ph}\) also saturates to a finite value, much smaller than one for the considered values of \(g\). (c) Electron density \(n_j=\langle c_j^\dagger c_j\rangle\) for \(L=200\) shows the robustness of topological edge states away from dimerized limit \(t_e=t_o/2=0.5\), for a system at half-filling minus one electron. The states can be seen as charge deficit at the two ends of the chain, which is insensitive to the strength of the light-matter coupling within numerical precision. Inset shows ED spectral gap \(\Delta\) for (\(L\leq 12\)) exhibiting the expected exponential decay with system size.
    \label{fig:spectrum}
    }
\end{figure*}

We address these issues through analytical and numerical methods that reveal a detailed account of the entanglement features, spectral properties, and edge states of SSH-cQED low energy states.
Departing from previous mean-field studies \cite{Dmytruk2022,perez2022topology},
we employ density-matrix renormalization group (DMRG) as a non-perturbative method to extract the structure of entanglement between light and electrons of the SSH chain as a function of light-matter coupling and of system size. 
Our DMRG analysis shows that, while LMI induces an expected increase in entanglement entropy due to LMI, this entanglement contribution \textit{saturates} with system size despite the non-locality of light-matter interactions. 
This behavior is associated with a many-body state characterized by dressed photon and electronic states which, nevertheless, preserves the area-law scaling of entanglement and the topological edge states, which are the central results of this work. 

Our DMRG analysis establishes an interesting correlation between EE saturation and the diamagnetic response of the ground state. This diamagnetic response is a non-perturbative feature that highlights the importance of gauge invariance in describing the interaction of Bloch electrons with quantum light \cite{li2020electromagnetic,dmytruk2021gauge}. While gauge invariant diamagnetic effects have been linked with stability against superradiant phase transitions \cite{Viehmann_Superradiant_2011, Rzaifmmode_Phase_1975, Slyusarev1979, Bamba_Stability_2014, Andolina_Cavity_2019}, 
the link between diamagnetism and quantum entanglement is a new aspect of LMIs that this work uncovers.

Furthermore, we corroborate the DMRG results by performing exact diagonalization (ED) and by identifying a closed Krylov subspace where light-matter entanglement can be efficiently described when the number of virtual photons in the ground state is small. The Krylov subspace is generated from the decoupled state of matter and photons by action of a composite operator involving creation and annihilation photons operators and a many-body fermionic current operator. This subspace spans the many-body ground state characterizing short-range entanglement of light and matter degrees of freedom observed in DMRG.
\\

\noindent
\textit{Model - }
Adopting the Coulomb gauge \cite{cohen1997photons},
we consider a half-filled chain of spinless fermions with $L = 2N$ sites described by creation (annihilation) operators
$c^{\dagger}_j$ ($c_j$) coupled to a single cavity transverse photonic mode of frequency $\omega$ represented by canonical bosonic operators $a^{\dagger}$ and $a$, which is described by the Hamiltonian
\begin{equation}
\label{eq: H cavity-matter Peierls}
H = 
\sum_{j} t_{j} e^{i\,\frac{e}{\hbar}\ell\mathcal{A}_{0}\,(a + a^{\dagger})}c^{\dagger}_{j}\,c_{j+1} + \textrm{H.c.} 
+
\hbar\,\omega\,a^{\dagger}\,a
\,,
\end{equation}
where the LMI is encoded via the Peierls substitution, and interactions
$V_{ij}c^{\dagger}_{i}c_{i}c^{\dagger}_{j}c_{j}\,$ mediated by the longitudinal component of the gauge field are disregarded since the fermionic matter is weakly correlated.
Nearest-neighbor intra- (inter-) unit cell real hopping amplitudes are, respectively,
$t_{2j} = t_e$ and $t_{2j+1} = t_o$ (see Fig.~\ref{fig: SSH-cavity}), with $t_o > t_e$ giving the topological SSH phase.  
The distance between neighboring sites is $\ell$, $e$ is the electron charge, $\hbar$ is Planck's constant divided by $2\pi$. 
The vector potential, polarized along the chain direction, has an amplitude
$
\mathcal{A}_{0} = 
\sqrt{
\frac{\hbar}
{2 \omega V \epsilon}
}
$
where $V$ is the cavity volume and $\epsilon$ is the dielectric constant. We fix the cross-sectional area of the cavity and consider the same length $L$ for the cavity and the chain (Fig.~ \ref{fig: SSH-cavity}). As such, the Peierls phase in \eqref{eq: H cavity-matter Peierls} $\frac{e \, \ell\,\mathcal{A}_{0}}{\hbar} \equiv g/\sqrt{L}$ explicitly encodes size variations of the chain size and the dimensionless strength \(g\) of LMI, which shall be varied from weak- to ultra-strong coupling regimes. We measure length in units of $\ell$ and regard $L$ as dimensionless.
\\ 

\begin{figure*}[t]
    \centering
    \includegraphics[width=0.99\linewidth]{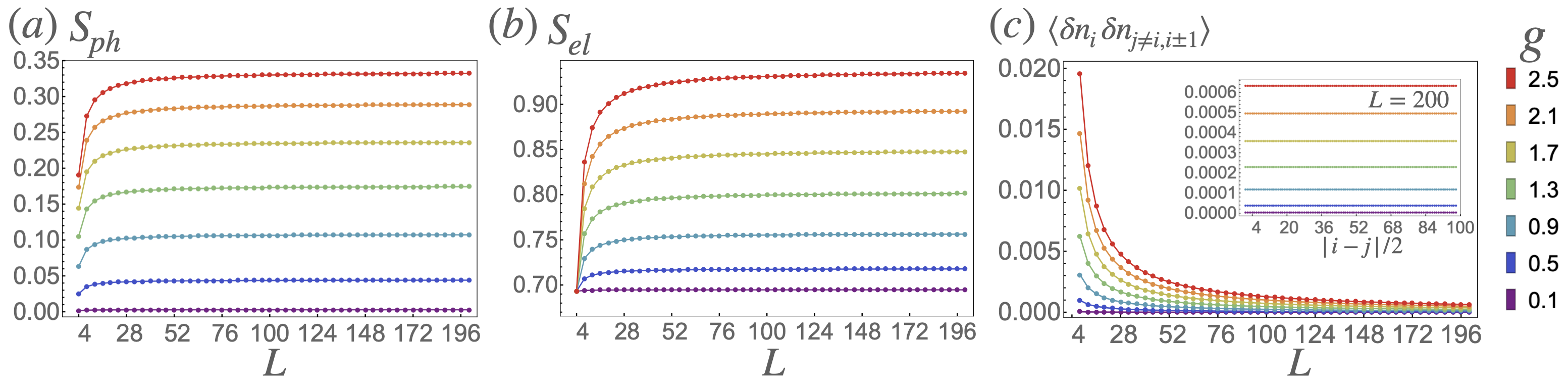}
    \caption{Entanglement entropy and density correlations of the cavity SSH system in the dimerized limit (\(t_e=0\)) found in DMRG for the same parameters as in Fig. \ref{fig:spectrum}. (a-b) The entanglement entropies between the photon and the fermions, \(S_{ph}\), and between the right half of the chain and the rest of the system, \(S_{el}\), versus system size. Both exhibit area law behavior in the thermodynamic limit. In the absence of light-matter coupling, \(S_{el}=\ln 2\) due to the non-trivial topology of the SSH chain. (c) Charge density correlation function \(\langle\delta n_i \delta n_j\rangle\) for \(i\) and \(j\) not belonging to the same dimer; as shown in the inset, the charge correlation function is independent of \(i\) and \(j\) (however, the sign alternates as \(\langle\delta n_i \delta n_j\rangle\propto (-1)^{i+j}\); not shown in the figure for clarity). The correlation length is infinite for fixed $L$, with a magnitude that decays as \(1/L\).
    \label{fig:EE}
    }
\end{figure*}

\noindent
\textit{Numerical Analysis - }
Using TeNPy \cite{tenpy}, we conducted a DMRG study of the model~Eq. \ref{eq: H cavity-matter Peierls}, varying the system size up to \(L=200\) and capping the the number of photons to 100.
The DMRG results presented here are for the quasi-resonance condition
\(\hbar\omega=2t_o=2\), but we have verified that no qualitative changes incur upon varying $\omega$.
In this study, the dimensionless coupling $g$ is varied over a wide range between the weak coupling $(g \ll 1)$ and ultra-strong coupling $(g \gg 1)$  regimes, with DMRG analysis for $g \in [0.1, 2.5]$ presented here, and additional data shown in the supplementary materials (SM) \cite{SM}. 
As shown in Figs. \ref{fig:spectrum}(a)-(b) obtained for the dimerized limit \(t_o=1\) and \(t_e=0\), both the ground state energy change due to light-matter coupling \(\Delta E=E(g)-E(g=0)\) and the number of photons \(N_{ph}=\langle a^\dagger a\rangle\) plateau to a constant value as system size \(L\) increases, with the value of the plateau increasing with increasing \(g\).  
Notably, while the lowest-order
term in expansion of the Hamiltonian \eqref{eq: H cavity-matter Peierls},
$\delta H_1 = (g/\sqrt{L})(a+a^{\dagger})\,\sum_{j} i\,t_{j}\,c^{\dagger}_{j}\,c_{j+1} + \textrm{H.c.}$, yields a \textit{negative} Lamb shift \cite{cohen1997photons}, the $\Delta E > 0$ plateau in Fig.~\ref{fig:spectrum}(a)
highlights important diamagnetic effect, which contributes to the suppression of the number of ground state virtual photons as displayed in Fig.~\ref{fig:spectrum}(b).

An important finding of this work is the stability of the topological edge states, despite the non-locality of the cavity mode. This is seen explicitly in the dimerized limit \(t_e=0\) where the edge fermion operators \(c_0\) and  \(c_L\) remain \textit{decoupled} from the bulk owing to the gauge invariant form of the LMI.
We explicitly confirmed the stability of the edge states away from the dimerized limit by studying the case \(t_o=1\) and \(t_e=0.5\); the resulting electron density \(\langle n_j\rangle=\langle c_j^\dagger c_j\rangle\) along a chain of length \(L=200\) for a system at half-filling minus one electron is shown in Fig. \ref{fig:spectrum}(c), where the edge states are clearly seen as positive charge excess on both sides of the SSH chain. 
Note that the electron density for different \(g\) are identical within numerical precision, indicating that the light-matter coupling does not affect the electron density at all. 
Furthermore, an ED analysis [see inset of Fig.~\ref{fig:spectrum} (c)] confirms the existence of two quasi-degenerate lowest energy states separated by a gap \(\Delta\) that exponentially decreases increasing system size.

The robustness of the edge states in DMRG is verified deep in the ultra-coupling regime (\(g \sim 100\)) \cite{SM}. The same plateau behavior displayed in Fig.\ref{fig:spectrum} is observed for \(\Delta E\) and \(N_{ph}\) away from the dimerized limit, though the plateaus are not reached as quickly as in the dimerized limit. Moreover, the topologically trivial chain ($t_o < t_e$) displays similar bulk behavior for $\Delta E$, $N_{\textrm{ph}}$ and $n_j$, except for edge states that are not present in this case.

The spectral features described above are consistent with the structure of entanglement in the many-body ground state found in the DMRG simulation. This can be seen in Figs. \ref{fig:EE} (a-b) for the dimerized limit, showing the system size scaling of the entanglement entropy (EE) between the photon and the chain of electrons \(S_{ph}=-\text{Tr}\left[\rho_{ph}\ln \rho_{ph}\right]\), and the EE of half of the electron chain (with the entanglement cut across the strong bond) with the rest of the chain and the photon \(S_{el}=-\text{Tr}\left[\rho_{el}\ln \rho_{el}\right]\), where $\rho_{ph}$ and $\rho_{el}$ are the corresponding reduced density matrices.
The observed \textit{area law} scaling of EE in the presence of light-matter coupling is another key result of this work. The behavior of $S_{ph}$ in Fig. \ref{fig:EE} (a) is similar to the saturation of the virtual photons $N_{ph}$ in Fig.~\ref{fig:spectrum} (b), as further discussed in Eq.\eqref{Sph}. Moreover, LMI generates an additional contribution to the electronic EE $S_{el}$ in addition to $\ln 2$ for $g=0$ in the dimerized limit \(t_e=0\) (assuming \(L\) is divisible by four such that a non-trivial bond is cut in the bipartition), signifying that electronic states are dressed by the photon while the system remains short-ranged entangled. The stability of the short-range entangled SPT phase of the SSH chain is further confirmed by the double degeneracy of the entanglement spectrum of \(\rho_{el}\) \cite{pollmann2010entanglement,Fidkowski_Entanglement_2010,turner2011topological}.

The additional EE indicates the presence of interaction-induced correlations in the system: although, as seen in Fig. \ref{fig:spectrum} (c), the excess electron density \(\langle\delta n_j\rangle=\langle n_j\rangle-1/2\) is unchanged by the light-matter coupling, we find that it induces charge fluctuations \(\langle\delta n_i\delta n_j\rangle\) (for \(i\neq j, j\pm 1\)) with a characteristic \(1/L\) decay while having an infinite correlation length for \textit{fixed system size},
as seen in the constant value of\(\langle\delta n_i\delta n_j\rangle\) as a function of separation between the sites \(|i-j|\) at fixed system size \(L=200\) (the fluctuations change sign between even and odd values of \(|i-j|\); only even values are shown for clarity). This infinite correlation range persists to changes away from dimerized limit and stronger LMI, and is an important signature of the LMI.
In the dimerized limit, it follows from a permutation symmetry of Hamiltonian \eqref{eq: H cavity-matter Peierls} that exchanges pairs of dimers, resulting in many-body states where photons are entangled with gas of delocalized dimers that mediate such long-range correlation functions. However, despite the constancy of these correlations for fixed system size, the $1/L$ behavior indicates the absence of long-range order in the thermodynamic limit. 
\\

\noindent
\textit{Physical Interpretation of Low Energy States -}
Physical insight into numerical results can be gained by recasting Hamiltonian \eqref{eq: H cavity-matter Peierls} as
\begin{equation}
H= H_0 \cos \frac{g}{\sqrt{L}}(a+a^\dagger)+J \sin \frac{g}{\sqrt{L}}(a+a^\dagger)+\hbar\omega a^\dagger a
\end{equation}
where \(H_0 = \sum_j t_j c_j^\dagger c_{j+1}+\textrm{H.c.}\) 
and \(J=\sum_j it_j c_j^\dagger c_{j+1}+\textrm{H.c.}\) are the SSH Hamiltonian and electron current operators in the absence of LMI, respectively. 
Importantly, the hopping imbalance $t_o \neq t_e$ of the SSH chain is responsible for quantum fluctuations ($[J, H_{0}] \neq 0$) that manifest in matter sector of the ground state, as follows.
In the dimerized limit, many-body eigenstates of \(H_0\) are tensor products of dimer states \(|\psi_{l\pm}\rangle=(|0_{2l-1},1_{2l}\rangle\pm|1_{2l-1},0_{2l}\rangle)/\sqrt{2}\) expressed in occupation number basis, where \(l=1,\dots,N_b=(L-2)/2\) is a dimer index, \(N_b\) being the number of non-trivial bonds (recall that the \(j=0,L\) sites decouple from the Hamiltonian). Observe that \(J=\sum_l J_l = \sum_l i t_o c_{2j-1}^\dagger c_{2j} + \textrm{H.c.}\), where \(J_l\) act as ladder operators on the dimer states: \(J_l|\psi_{l\pm}\rangle=\pm it_o |\psi_{l\mp}\rangle\).
Let us therefore denote product states with dimers \(l_1,\dots,l_n\) in excited states \(|\psi_{l_j+}\rangle\) as \(|\Psi_{l_1,\dots,l_n}\rangle\). This allows us to identify the Krylov subspace of the ground state $H_0$ \(|\Psi^{(0)}\rangle=\bigotimes_{l}|\psi_{l-}\rangle\) by successive applications of \(J\).
This subspace is spanned by $|\Psi^{(0)}\rangle$ and the orthonormal states \(|\Psi^{(n)}\rangle\) describing \textit{uniform} superpositions of all \(\binom{N_b}{n}\) states with \(n\) excited dimers, similar to Dicke states \cite{Dicke54}:
\begin{equation}
\left|\Psi^{(n)}\right\rangle=\frac{1}{\sqrt{\binom{N_b}{n}
}}\sum_{0<l_1<l_2<\dots<l_n<N_b-1}\left|\Psi_{l_1,l_2,\dots,l_n}\right\rangle\,, \label{Psin}
\end{equation}
Importantly, the Hamiltonian can thus be brought into block-diagonal form with one of the blocks acting only on this \((N_b+1)\)-dimensional Krylov subspace.

At weak coupling, the ground state wavefunction is in the Krylov subspace of \(|\Psi^{(0)}\rangle\) and can thus be expressed as \(|\Xi\rangle=\sum_n \xi_n|\Psi^{(n)}\rangle|\gamma_n\rangle\) where \(|\gamma_n\rangle\) are photon states. Furthermore,
noting that \(N_{ph}\ll 1\) for small \(g\) as seen in Fig. \ref{fig:spectrum} (b), the ground state can be further approximated by capping the photon number to one, yielding an effective Hamiltonian
\begin{equation}
\label{eq: truncated H}
H'=H_0 \cos\left(g/\sqrt{L}\right) +  J \sin\left(g/\sqrt{L}\right)\sigma^x - \frac{\hbar\omega}{2}\, (\sigma^z-1)  
\end{equation}
where the Pauli matrices \(\sigma^j\) act on the two-state truncated photon Hilbert space. Upon projecting the Hamiltonian Eq. \eqref{eq: truncated H} onto the Krylov subspace, we obtain
\(|\gamma_n\rangle=|n \mod 2\rangle\) and the matrix equation
\begin{align}
    &\frac{E+(N_b-2n)t_o\cos \left(g/\sqrt{L}\right)+(1-(-1)^n)\hbar\omega/2}{t_o\sin \left(g/\sqrt{L}\right)}\xi_n=\nonumber\\
    &=i\sqrt{(n+1)(N_b-n)}\xi_{n+1}-i\sqrt{n(N_b-n+1)}\xi_{n-1}\label{Eq:evals}
\end{align}
for \(E\) and \(\xi_n\), which can be efficiently solved numerically for large system sizes. Remarkably, all observables including the energy, entanglement entropies and correlation functions obtained by solving \eqref{Eq:evals} are \textit{identical within numerical precision} to those obtained in DMRG with photon number restricted to at most one, confirming that \(|\Xi\rangle\) is an exact ground state of \(H'\).
In particular, the number of photons is 
$ N_{ph}=\sum_m |\xi_{2m+1}|^2 \label{Nph}$
and the photon EE takes an intuitive Gibbs form
\begin{equation}
S_{ph}=-(1-N_{ph})\ln (1-N_{ph})- N_{ph}\ln N_{ph}\label{Sph} 
\,,
\end{equation}    
since either a photon is created or not created in the weak coupling limit.
Eq.\eqref{Sph} then relates the area law for the photon EE seen in Fig. \ref{fig:EE} (a) and the saturation of the photon number \(N_{ph}\) in Fig.\ref{fig:spectrum} (b)
Analogous but lengthier expressions for 
\(S_{el}\) and \(\langle\delta n_i\delta n_j\rangle\) are given in the SM \cite{SM}.

We next perform perturbation theory of the Krylov theory to leading order in \(g\) by taking \(\xi_n=0\) for \(n>1\). Note that the first-order perturbation theory for \(H\) and \(H'\) are equivalent as only zero and one photon states appear in both cases. This yields
\begin{subequations}
\label{eq: 1st order Krylov}
    \begin{gather}
    N_{ph}=|\xi_1|^2=\frac{g^2}{4}\frac{1-2/L}{(1 + \hbar\omega/2t_o)^2}\,,
    \label{Nphpert}\\
    \Delta E =\left(1 + \frac{\hbar\omega}{2t_o}\right) \hbar\omega N_{ph}\,,
    \label{Epert}\\
    S_{el}=-(1-N_{ph})\ln\frac{1-N_{ph}}{2}-\frac{N_{ph}}{2}\ln\frac{N_{ph}}{8}\,,
    \label{Spert}\\
    \langle \delta n_{2l+1}\delta n_{2l'+1}\rangle=\frac{N_{ph}}{L-2}\,,
    \label{JJpert}
    \end{gather}
\end{subequations}
with \(l\neq l'\) in the last expression; for \(S_{el}\), we further assumed that \(L\) is large. We note that the charge fluctuations are related to current fluctuations in the space of dimer states, since \(\delta n_{2l+1}|\psi_{l\pm}\rangle= \mp i J_l/(2t_o)|\psi_{l\pm}\rangle\), and in particular \(\langle J_l J_{l'}\rangle=4t^2\langle \delta n_{2l+1}\delta n_{2l'+1}\rangle\) to leading order in perturbation theory. This is in agreement with the recent result in \cite{Passeti23} that found that \(S_{ph}=0\) iff \(\langle J_l J_{l'}\rangle=0\). However, the scaling analysis of the EE and the stability of the topological edge states, which are central results of this work, were not discussed in Ref.~ \cite{Passeti23}.
The excess in EE may therefore be measurable in transport experiments, which are sensitive to current fluctuations, for example in a setup proposed in \cite{KlichLevitov09}.

Eqs.~\eqref{Sph} and \eqref{eq: 1st order Krylov}) capture all of the qualitative aspects of the DMRG results shown in Fig.'s \ref{fig:spectrum} and \ref{fig:EE}: since \(\Delta E\), \(S_{ph}\), and \(S_{el}\) are determined by \(N_{ph}\), 
the saturation of \(N_{ph}\) in the thermodynamic limit dictates similar behavior for all other quantities.
The EE in particular follows the area law. Because the photon couples to fermions via the total current \(J\), \(N_{ph}\) scales linearly with 
$L$;
however, it is also proportional to the square of the light-matter coupling strength \(g^2/L\). It is the precise cancellation between these two factors that result in the saturation of \(N_{ph}\). Furthermore, DMRG simulations confirm that the qualitative features of the weak coupling regime described by \eqref{Sph} and \eqref{eq: 1st order Krylov}) persist all the way to the ultra-strong coupling \cite{SM}. In particular, while more photons are virtually created in the ground state at stronger LMI,
correlation functions $\langle \delta n_i \delta n_j \rangle$ in the dimerized limit display long range behavior consistent with the ground state being spanned by a uniform superposition of dimers belonging to the Krylov subspace of \(|\Psi^{(0)}\rangle\). The identification of this subspace strongly suggests a remarkable connection between light-matter entanglement and Hilbert space fragmentation \cite{Znidaric2013coexistence, iadecola2019exact, Khemani2020localization, Moudgalya_scars_2020, Sala2020ergodicity}, a 
scenario worthy of further examination.

In summary, we have characterized the effects of light-matter interaction on the SSH-cQED low energy states, employing numerical methods (DMRG, ED) and a low-dimensional Krylov subspace effective theory. We have established the stability of the topological edge states despite long-range correlations induced by the interaction of electrons with a uniformly extended cavity mode. This work highlights how gauge invariance, diamagnetic effects, and electron-photon entanglement give rise to an area law scaling of the entanglement entropy despite the non-locality of light-matter interactions. Extending this approach to higher dimensional topological phases in cavity material systems offers a promising path to classify novel light-matter hybrid states of matter. We leave such matters for future investigation.
\\

\noindent
\textit{Acknowledgements.} We thank Claudio Chamon, Raman Sohal, and the participants of the Quantum Science Gordon Research Conference ``Many-Body Quantum Systems: From Quantum Computing and Simulation to Metrology and Coherent Light-Matter Hybrids" for useful discussions. This research was supported by the U.S. Department of Energy, Office of Science, Basic Energy Sciences, under Award DE-SC0023327 (D.S. and L.H.S.), and the National Science Foundation, under Award DMR-2132591 (M.C.). DMRG simulations were performed on the mobius cluster at the University of Pennsylvania.

\bibliographystyle{apsrev4-1}
\bibliography{bibliography}

s

\pagebreak
\widetext
\pagebreak
\begin{center}
\textbf{\large Supplementary Material for Entanglement and Topology in Su-Schrieffer-Heeger Cavity Quantum Electrodynamics}
\end{center}
\setcounter{equation}{0}
\setcounter{figure}{0}
\setcounter{table}{0}
\setcounter{page}{1}
\makeatletter
\renewcommand{\theequation}{S\arabic{equation}}
\renewcommand{\thefigure}{S\arabic{figure}}
\renewcommand{\bibnumfmt}[1]{[S#1]}
\renewcommand{\citenumfont}[1]{S#1}

\section{Additional DMRG Data}

\begin{figure*}[b]
    \centering
    \includegraphics[width=0.99\linewidth]{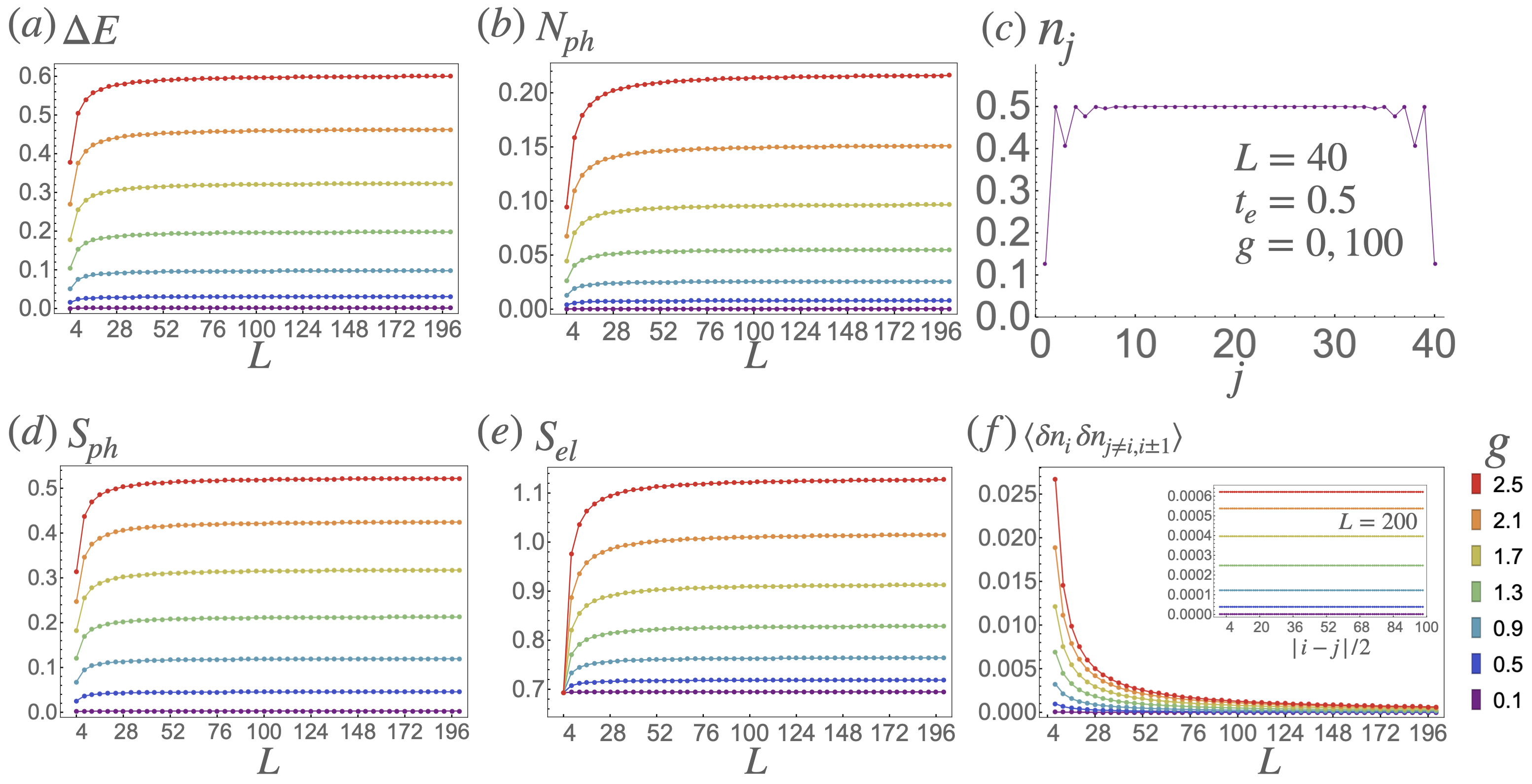}
    \caption{(a-c) Same as Fig. 2, (d-f) same as Fig. 3, but with the number of photons capped to at most one. Both DMRG and expression found using solutions of Eq. (5) in the Krylov subspace produce identical plots (a-b) and (d-f), within numerical precision. For plot (c) showing edge states in the local electron density we took \(L=40\) and \(g=0\) and \(100\), which have equal electron densities within numerical precision.
    \label{fig:capped}
    }
\end{figure*}

\begin{figure*}[b]
    \centering
    \includegraphics[width=0.99\linewidth]{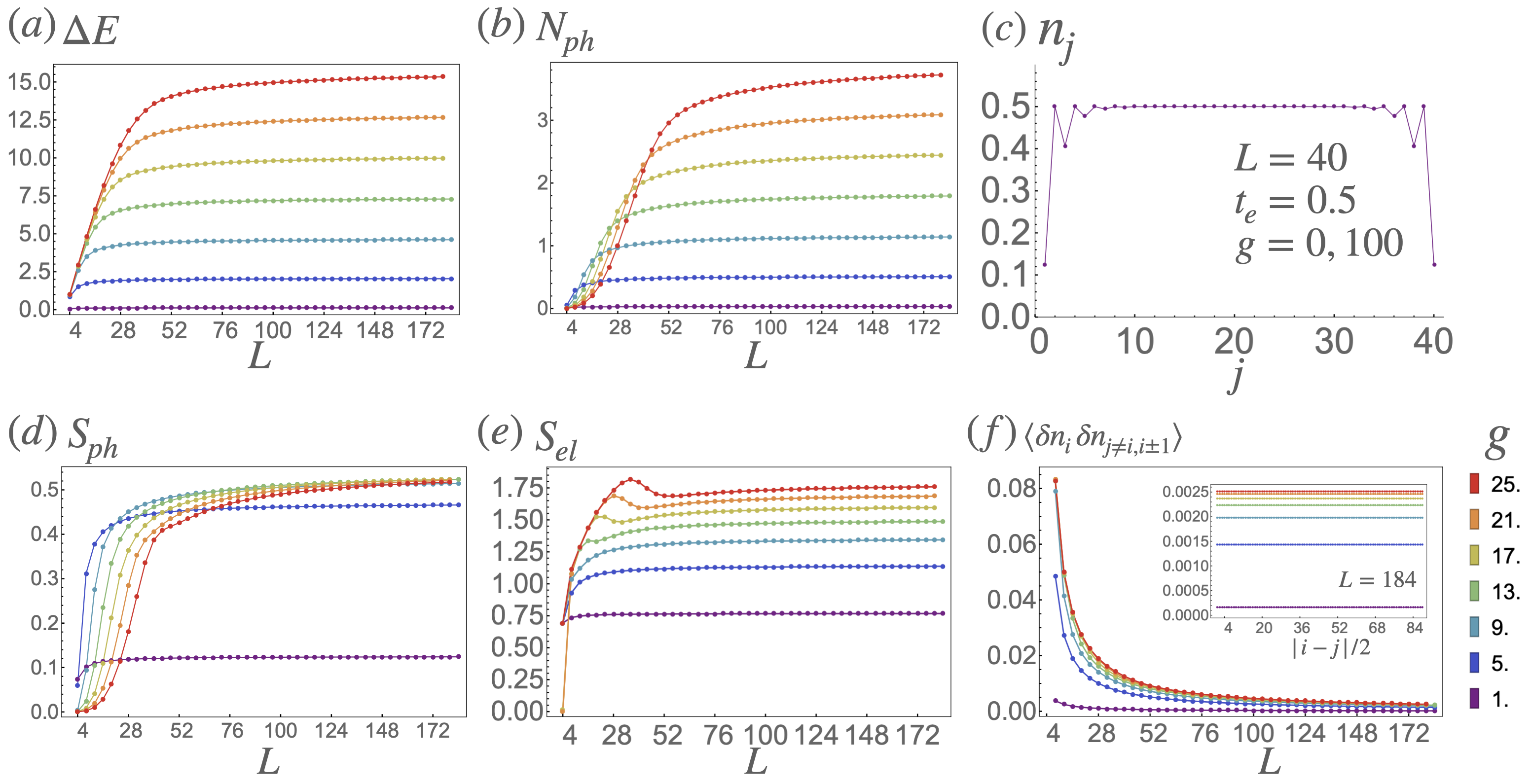}
    \caption{(a-c) Same as Fig. 2, (d-f) same as Fig. 3, but with the \(g\) increased by a factor of ten. For plot (c) showing edge states in the local electron density we took \(L=40\) and \(g=0\) and \(100\), which have equal electron densities within numerical precision. We note that in subfigure (e) the \(S_{el}\approx 0\) points at \(L=4\) appear to be numerical aberrations, as varying \(g\) slightly above \(25\) again gives \(S_{el}=\ln 2\); we speculate that this may be due to states from two different Krylov subspaces being close in energy around those values of \(g\).
    \label{fig:strong}
    }
\end{figure*}

Here we present some plots for additional parameter values. Fig. \ref{fig:capped} shows the same plots as in Figs. 2 and 3, but with the number of photons capped to at most one, i.e. for the Hamiltonian in Eq. (4). Both DMRG and expression found using solutions of Eq. (5) in the Krylov subspace produce identical plots (a-b) and (d-f), within numerical precision. We note that the number of photons, and consequently other quantities, is somewhat overestimated for larger values of \(g\). 
Fig. \ref{fig:strong} is also the same as Figs. 2 and 3, but with \(g\) increased by a factor of ten. Some slight changes can be seen as the average number of photons exceeds \(1\), in particular \(N_{ph}\) develops an inflection point and is no longer always increasing in \(g\) at fixed small \(L\). In the thermodynamic limit, however, the qualitative behavior is the same as at small \(g\).

\section{Calculations of  EE and Correlations from Solutions of Eq. (5)}

Here we calculate EE for arbitrary connected bipartitions of the cavity-SSH system in the dimerized limit and with the number of photons capped to one, given the solution of Eq. (5) \(|\Xi\rangle=\sum_n \xi_n |\Psi^{(n)}\rangle\).
Since \(\left|\Psi^{(n)}\right\rangle\) are uniform superpositions of \(\left|\Psi_{l_1,l_2,\dots,l_n}\right\rangle\) that are all orthonormal states, the singular values for an arbitrary cut of the chain are relatively easy to find. We need to perform a a Schmidt decomposition over the bonds and the photon. First let us assume the cut happens on a non-trivial bond, with \(n_L\) bonds/dimers fully on the left of the cut and \(n_R\) fully on the right, so that \(N_b=n_L+n_R+1\) (\(n_L=n_R=\frac{N_b-1}{2}\) for a half-chain cut). To carry out the Schmidt decomposition we first need to write the eigenstate as
\begin{align}
    |\Xi\rangle&=\sum_{m=0}^{n_L} \sum_{n=0}^{n_R} \psi_{mn\pm}|\Psi^{(m)}\rangle_L |\psi_{\pm}\rangle_{C} |\Psi^{(n)}\rangle_R\left|\left(n+m+\frac{1\pm1}{2}\right)\mod 2\right\rangle=\nonumber\\
    &=\sum_{l=0,1}\sum_{m=0}^{n_L} \sum_{m=0}^{n_R} \alpha_{m,2n+l}|\Psi^{(m)}\rangle_L |\psi_{(-1)^{m+n+l}}\rangle_{C} |\Psi^{(n)}\rangle_R\left|l\right\rangle
\end{align}
where
\[|\Psi^{(m)}\rangle_{L/R}=\frac{1}{\sqrt{\binom{n_{L/R}}{m}
}}\sum_{j_1<j_2<\dots<j_m}\left|\Psi_{j_1,j_2,\dots,j_m}\right\rangle_{L/R}\]
with \(\left|\Psi_{j_1,j_2,\dots,j_m}\right\rangle_{L/R}\) defined the same way as \(\left|\Psi_{j_1,j_2,\dots,j_m}\right\rangle\) but with the dimer indices restricted to the left or right sides, \(j_1,\dots,j_m\in L\) or \(R\). \(\alpha\) is consequently a \((n_L+1)\times 2(n_R+1)\) matrix with elements \(\alpha_{m,2n+l}=\psi_{mn,(-1)^{m+n+l}}\), and we have
\begin{align}
\psi_{mn+}&=\xi_{m+n+1}\sqrt{\binom{n_L}{m}\binom{n_R}{n}\left/\binom{N_b}{n+m+1}\right.}\nonumber\\
\psi_{mn-}&=\xi_{m+n}\sqrt{\binom{n_L}{m}\binom{n_R}{n}\left/\binom{N_b}{n+m}\right.}
\end{align}
which we note means that \(|\psi_{mn\pm}|^2\) asymptotically follow the hypergeometric distribution. 

Since the central dimer is split into left and right parts as well
\[|\psi_{\pm}\rangle_{C}=\frac{1}{\sqrt{2}}\left(|0\rangle_L|1\rangle_R\pm|1\rangle_L|0\rangle_R\right)\]
the matrices we actually want to carry out the singular value decomposition on are \(\alpha/\sqrt{2}\) and \(\alpha'/\sqrt{2}\) with \(\alpha'_{m,2n+l}=(-1)^{m+n+l}\alpha_{m,2n+l}\). Note that \(\alpha'\) and \(\alpha\) have the same singular values (since we can take \(|\psi_\pm\rangle\rightarrow\pm|\psi_\pm\rangle\)).
There are thus in general \(2\min[n_L+1,2(n_R+1)]\) two-fold degenerate singular values \(\lambda_m\) of \(\alpha\), and the entanglement entropy is given by
\[S_{VN}=-2\sum_{m}\lambda_m^2 \ln \lambda_m^2\]
with \(S_{el}=S_{VN}\) for \(n_L=n_R=(N_b-1)/2\) (which assumes \(N_b\) is odd).


The calculation is essentially the same when the chain is cut at a trivial bond, and in fact a bit simpler, since now the eigenstate can be written simply as
\begin{align}
|\Xi\rangle&=\sum_{m=0}^{n_L} \sum_{n=0}^{n_R} \psi_{mn}|\Psi^{(m)}\rangle_L |\Psi^{(n)}\rangle_R\left|\left(n+m\right)\mod 2\right\rangle=\nonumber\\
&=\sum_{l=0,1}\sum_{m=0}^{n_L} \sum_{n=0}^{n_R} \alpha_{m,2n+l}|\Psi^{(m)}\rangle_L |\Psi^{(n)}\rangle_R\left|l\right\rangle
\end{align}
with \(\alpha_{m,2n+l}=\psi_{mn}\delta_{l,(m+n)\mod 2}\) and
\[\psi_{mn}=\xi_{m+n}\sqrt{\binom{n_L}{m}\binom{n_R}{n}\left/\binom{N_b}{n+m}\right.}\]
but now with \(n_L+n_R=N_b\). In this case the singular values are not doubly degenerate. The entanglement entropy is computed the same way as before.
As a special case, we consider the entanglement entropy of the photon with the whole chain, i.e. \(n_L=N_b\) and \(n_R=0\). We then have \(n=0\), \(m=0,\dots,N_b\), so \(\psi_{m,0}=\xi_m\).
To find the singular values, we can compute \(\alpha \alpha^\dagger\), which in this case is a \(2\times 2\) matrix:
\[[\alpha \alpha^\dagger]_{l,l'}=\sum_{m}\alpha_{m,l}\alpha^*_{m,l'}=\sum_{m}(|\xi_{2m}|^2\delta_{l0}+|\xi_{2m+1}|^2\delta_{l1})\delta_{ll'}\]
so the squares of the singular values are simply \(\lambda_0^2=\sum_{m}|\xi_{2m}|^2\) and \(\lambda_1^2=\sum_{m}|\xi_{2m+1}|^2\). Note that \(N_{ph}=\langle a^\dagger a\rangle=\lambda_1^2\)

\subsection{Correlation Functions}

Here we compute the correlation function \(\langle\delta n_i \delta n_j\rangle\), with \(\delta n_i=c_i^\dagger c_i-1/2\). To do this, we first observe that when acting on dimer states,
\[\delta n_{2i+1}|\psi_{i\pm}\rangle=\delta n_{2i+1} \left(|0_{2i+1},1_{2i+2}\rangle\pm |1_{2i+1},0_{2i+2}\rangle\right)/\sqrt{2}=|\psi_{i\mp}\rangle/2\]
and similarly
\[\delta n_{2i+2}|\psi_{i\pm}\rangle=-|\psi_{i\mp}\rangle/2\]
so that acting on the dimer states \(\delta n_{2i+1}\equiv \mp i J_i/(2t)\) where \(J_i=it c_{2i+1}^\dagger c_{2i+2}+h.c.\).

From this, we can deduce the action of \(\delta n_j\) on the \(|\Psi^{(n)}\rangle\) states:
\[2\delta n_{2i+1} |\Psi^{(n)}\rangle=\sqrt{1-\frac{n}{N_b}}|\Psi^{(n+1)}_i\rangle+\sqrt{\frac{n}{N_b}}|\Psi^{(n-1)}_{\cancel{i}}\rangle\]
where we defined
\begin{align}
    |\Psi^{(n)}_i\rangle&=\frac{1}{\sqrt{\binom{N_b-1}{n-1}}}\sum_{j_{n-1}\neq i}\sum_{i\neq j_{n-2}< j_{n-1}}\dots \sum_{i\neq j_1<j_2}|\Psi_{j_1,\dots,j_{n-1},i}\rangle\nonumber\\
    |\Psi^{(n)}_{\cancel{i}}\rangle&=\frac{1}{\sqrt{\binom{N_b-1}{n}}}\sum_{j_{n}\neq i}\sum_{i\neq j_{n-1}< j_n}\dots \sum_{i\neq j_1<j_2}|\Psi_{j_1,\dots,j_{n}}\rangle
\end{align}
Thus \(|\Psi^{(n+1)}_i\rangle\) comes from applying \(\delta n_{2i+1}\) to dimer states in \(|\Psi^{(n)}\rangle\) in which the \(i^{th}\) dimer is not excited (and becomes excited), while \(|\Psi^{(n)}_{\cancel{i}}\rangle\) comes from the dimer states in which the \(i^{th}\) dimer is excited (which becomes unexcited).

We then need to compute the inner products for the \(|\Psi^{(n)}_i\rangle\) and \(|\Psi^{(n)}_{\cancel{i}}\rangle\) states. First, we have
\[\langle\Psi^{(m)}_i|\Psi^{(n)}_j\rangle=\delta_{ij}+(1-\delta_{ij})\delta_{mn}\frac{n-1}{N_b-1}\,.\]
The second coefficient comes from the fact that the inner product of the dimer states \(|\Psi_{i_1,\dots,i_{m-1},i}\rangle\) in \(|\Psi^{(m)}_i\rangle\) and \(|\Psi_{j_1,\dots,j_{m-1},j}\rangle\) in \(|\Psi^{(n)}_j\rangle\) are non-zero only for \(i\neq j\)  when one of \(i_l=j\) (there is then exactly one \(|\Psi_{j_1,\dots,j_{m-1},j}\rangle\) with the same excited dimer configuration). There are \(\binom{N_b-2}{n-2}\) such \(|\Psi_{i_1,\dots,i_{m-1},i}\rangle\), while the normalization is \(1/\binom{N_b-1}{n-1}\). By similar reasoning, we find
\[\langle\Psi^{(m)}_{\cancel{i}}|\Psi^{(n)}_{\cancel{j}}\rangle=\delta_{ij}+(1-\delta_{ij})\delta_{mn}\frac{N_b-n-1}{N_b-1}\]
and
\[\langle\Psi^{(m)}_{i}|\Psi^{(n)}_{\cancel{j}}\rangle=(1-\delta_{ij})\delta_{mn}\frac{\sqrt{n(N_b-n)}}{N_b-1}\,.\]

Using these expressions, we compute
\[\langle \delta n_{2i+1}\delta n_{2j+1}\rangle=\langle \Xi|\delta n_{2i+1}\delta n_{2j+1}|\Xi\rangle=\sum_{nm}\xi_m^*\xi_n \langle \Psi^{(m)}|\delta n_{2i+1}\delta n_{2j+1}|\Psi^{(n)}\rangle\]
for \(i\neq j\) (when \(i=j\) we simply get \(1/4\)). The result is
\[\langle \delta n_{2i+1}\delta n_{2j+1}\rangle=\frac{1}{4N_b(N_b-1)}\sum_{n=0}^{N_b}\left(2n(N_b-n)|\xi_{n}|^2+2\text{Re}\left[\xi_{n-2}^*\xi_n\sqrt{n(n-1)(N_b-n+2)(N_b-n+1)}\right]\right)\]
(for concreteness, we take \(\xi_{-2}=\xi_{-1}=0\)). We also observe that the current-current correlations have a similar form:
\[\langle J_i J_{j\neq i}\rangle=\frac{t^2}{N_b(N_b-1)}\sum_{n=0}^{N_b}\left(2n(N_b-n)|\xi_{n}|^2-2\text{Re}\left[\xi_{n-2}^*\xi_n\sqrt{n(n-1)(N_b-n+2)(N_b-n+1)}\right]\right)\]
with the main difference being the sign in the second term.

\subsection{Perturbation Theory}

Assuming \(\sin \beta\ll1\) (i.e. \(\beta\ll1\)), in leading order of perturbation theory we can keep \(\xi_n\) only up to \(n=1\). The Hamiltonian projected onto this restricted space reads simply
\[H'\rightarrow \left(\begin{array}{cc}
    -N_b t \cos \beta & -i \sqrt{N_b} t \sin \beta \\
    i \sqrt{N_b} t \sin \beta & \hbar\omega - (N_b-2) t \cos \beta
\end{array}\right)\]
From this, we find that for the ground state
\begin{align}
 E&=\frac{1}{2} \left(\hbar\omega - 2 (L-1) t \cos \beta - \sqrt{
 (\hbar\omega + 2 t \cos g)^2+4 L t^2 \sin^2 \beta}\right)\approx-N_b t + \frac{t \hbar\omega \beta^2 N_b}{4 t + 2 \hbar\omega}\nonumber\\
 (\xi_0,\xi_1)&\approx\left(1 - \frac{N_b t^2 \beta^2}{2(2 t + \hbar\omega)^2},-i\frac{\sqrt{N_b} t \beta}{2 t + \hbar\omega}\right)
\end{align}
Eq. (7) in the main text follows with \(\beta=g/\sqrt{L}\).

\end{document}